\documentclass[aps, prd, onecolumn, tightenlines, notitlepage, superscriptaddress, nofootinbib, preprintnumbers, floatfix,showkeys,11pt,altaffilletter]{revtex4-2}

\usepackage[T1]{fontenc}
\usepackage[utf8]{inputenc}
\usepackage{lmodern}

\usepackage{amsmath}
\usepackage{amssymb}
\usepackage{amsfonts}
\usepackage{amsthm}
\usepackage{bm}
\usepackage{slashed}
\usepackage{tabularx}
\usepackage{array}
\usepackage{comment}
\renewcommand{\arraystretch}{1.3}

\usepackage{graphicx}
\usepackage{xcolor}
\usepackage{subcaption}

\usepackage{booktabs}
\usepackage{multirow}
\usepackage{float}

\usepackage[
colorlinks=true,
linkcolor=blue,
citecolor=blue,
urlcolor=blue
]{hyperref}

\usepackage{tikz}
\usepackage[compat=1.1.0]{tikz-feynman}

\usepackage{siunitx}

\begin{document}

\title{Fermiophobic Z' Bosons in 3-4-1 Models: Classification  and Collider Constraints}

\author{Eduard Suarez}
\affiliation{Departamento de física, Universidad Nacional de Trujillo}

\author{Yithsbey Giraldo}
\affiliation{Departamento de F\'\i sica, Universidad de Nari\~no, A.A. 1175,  San Juan de Pasto, Colombia.}

\author{Eduardo Rojas}
\affiliation{Departamento de F\'\i sica, Universidad de Nari\~no, A.A. 1175,  San Juan de Pasto, Colombia.}

\footnotetext{Electronic address:
esuarezar@gmail.com}
\footnotetext{Electronic address:
eduro4000@gmail.com}
\footnotetext{Electronic address:
yithsbey@gmail.com}

\date{\today}

\begin{abstract}
We investigate the fermion spectra and collider phenomenology of models based on the \(SU(3)_C \otimes SU(4)_L \otimes U(1)_X\) gauge symmetry, focusing on two realizations of the electric charge operator characterized by \(b=c=1\) and \(b=1,\ c=-2\). For each configuration, we classify anomaly-free fermion sets and identify five minimal spectra with at least three SM-like quark and lepton families for each configuration, distinguishing algebraically irreducible structures from reducible spectra that remain minimal under the three-family requirement. We derive the chiral couplings of the additional neutral gauge boson \(Z'\) to Standard Model fermions and show that distinct anomaly-free fermion spectra can lead to identical low-energy neutral-current interactions. We then investigate the corresponding collider constraints using the ATLAS dilepton resonance search at \(\sqrt{s}=13\) TeV with an integrated luminosity of \(139\,\mathrm{fb}^{-1}\), assuming negligible \(Z-Z'\) mixing and kinematically inaccessible exotic fermions. A particularly interesting result emerges for the \(b=c=1\) configuration, where all \(Z'\) couplings to Standard Model fermions are proportional to \(\sin\psi\), with \(\psi\) denoting the additional neutral gauge-boson mixing angle. Consequently, these couplings vanish simultaneously at \(\psi=n\pi\), giving rise to a fermiophobic limit in which conventional Drell–Yan dilepton constraints disappear. In contrast, the \(b=1,\ c=-2\) configuration does not exhibit a universal fermiophobic limit. Our results demonstrate that collider constraints on additional neutral gauge bosons depend sensitively on the fermion embeddings and neutral gauge-boson mixing structure, and suggest the possibility of comparatively light \(Z'\) bosons in suitable realizations of extended electroweak gauge theories.
\end{abstract}

\maketitle



\section{Introduction}

The Standard Model (SM) of particle physics offers a highly precise description of fundamental interactions, yet it leaves several structural questions unanswered. Prominent among these is the ``flavour puzzle''---the unexplained replication of three fermion generations and the vast hierarchy observed in their mass spectrum. A compelling framework that addresses this issue structurally involves extending the SM electroweak gauge group $SU(2)_L \times U(1)_Y$ to a larger symmetry. A classic historical step was the $SU(3)_C \times SU(3)_L \times U(1)_X$ model (the 3-3-1 model)~\cite{Pisano:1992bxx,Frampton:1992wt}, which was subsequently generalized to its natural higher-dimensional counterpart, the $SU(3)_C \times SU(4)_L \times U(1)_X$ symmetry, commonly referred to as the 3-4-1 model~\cite{Pisano:1994tf,Foot:1994ym}.

The primary motivation for introducing the 3-4-1 class of models is that gauge anomaly cancellation requires the number of fermion generations to be a multiple of the number of quark colours ($N_c = 3$). By embedding the three families non-trivially under different representations of the larger gauge group, the cancellation of triangle gauge anomalies enforces three generations, directly linking the macroscopic number of families to the microscopic structure of quantum chromodynamics (QCD). Furthermore, the embedding of SM fermion families into $SU(4)_L$ quadruplets allows for a unified treatment of right-handed neutrinos within the same multiplet as the left-handed leptons~\cite{Pisano:1994tf}, providing a natural landscape for neutrino mass generation mechanisms.

Over the years, various realizations of the 3-4-1 symmetry have been comprehensively explored in the literature. Foundational classification and structural formulations can be traced to the works exploring both exotic and non-exotic electric charge assignments~\cite{Nisperuza:2009xm,Villada:2009iu}. A particularly notable variation is the compact 3-4-1 model at the TeV scale~\cite{Dias:2013kma}, which minimizes the scalar content required for symmetry breaking and utilizes non-renormalizable effective operators to explain neutrino masses. Additionally, alternative mass generation schemes via dimension-five effective operators have been extensively characterized~\cite{Palcu:2011tw,Palacio:2016mam}, establishing a solid groundwork for the scalar sector configurations.

Testing these frameworks requires confronting them with precise low-energy observations and direct collider searches at the Large Hadron Collider (LHC). The presence of an enlarged gauge sector introduces a rich spectrum of new heavy neutral gauge bosons ($Z^\prime$ and $Z^{\prime\prime}$) and exotic charged or neutral fermions. Low-energy experiments, including precision measurements at the $Z$-pole, atomic parity violation (APV), and neutral meson mixing ($K^0-\bar{K}^0$ and $B_d^0-\bar{B}_d^0$ oscillations), impose severe lower bounds on the new symmetry-breaking scale, typically pushing the masses of the extra gauge bosons above several TeVs~\cite{Villada:2009iu,Cogollo:2014tra}.

In recent years, attention has increasingly focused on the 3-4-1 models as an explanation for persistent anomalies in the lepton sector, such as the anomalous magnetic moment of the muon $(g-2)_\mu$ and lepton flavour-violating (LFV) decays like $e_b \to e_a \gamma$. Recent literature has successfully integrated these bounds by supplementing the gauge architecture with minimal or inverse seesaw mechanisms~\cite{Thao:2023gvs,Nha:2026pvq}.

In this paper, we systematically investigate anomaly-free fermion spectra for two representative realizations of the $SU(3)_C \times SU(4)_L \times U(1)_X$ gauge symmetry and derive the corresponding collider constraints. 

The remainder of this article is organized as follows. In Sec. II, we introduce the general structure of the \(SU(3)_C \otimes SU(4)_L \otimes U(1)_X\) gauge symmetry and define the electric charge operator. In Secs. III and IV, we present the fermion multiplets and their contributions to gauge anomalies for the configurations \(b=c=1\) and \(b=1,\ c=-2\), respectively. In Sec. V, we classify the irreducible anomaly-free fermion sets and construct the corresponding three-family models. In Sec. VI, we derive the chiral couplings of the additional neutral gauge boson \(Z'\) to Standard Model fermions. In Sec. VII, we investigate the collider constraints from ATLAS dilepton resonance searches, with particular emphasis on the fermiophobic limit arising in the \(b=c=1\) configuration. Finally, our conclusions are presented in Sec. VIII. The explicit chiral \(Z'\) charges for the different fermion representations are collected in Appendix A.

\section{The \texorpdfstring{$SU(3)\otimes SU(4)_L \otimes U(1)_X$}{} Gauge Group}
Since any 3-4-1 model must reproduce the Standard Model (SM) at low energies, it is necessary to identify the electromagnetic subgroup $U(1)_Q$ from the extended electroweak gauge symmetry
$SU(4)_L \otimes U(1)_X$. Equivalently, one must construct the electric charge operator within this extended gauge structure. In the most general 3-4-1 model, the electric charge operator is defined as
\begin{equation}\label{qem}
Q=aT_{3L}+\frac{b}{\sqrt{3}}T_{8L}
+\frac{c}{\sqrt{6}}T_{15L}
+XI_4,
\end{equation}
where $T_{iL}=\lambda_{iL}/2$ ($i=1,2,\ldots,15$), with $\lambda_{iL}$ denoting the generalized Gell--Mann matrices of $SU(4)_L$, normalized according to
$\mathrm{Tr}(\lambda_i\lambda_j)=2\delta_{ij}$. Here,
$I_4=\mathrm{diag}(1,1,1,1)$ is the identity matrix $4\times4$, 
$X$ is the quantum number associated with the gauge group $U(1)_X$, whose assignments are determined by requiring that the multiplets contain the SM doublets, and $b$ and $c$ are free parameters that can be fixed according to the electric charge assignments of the fermion content.
Consequently, an infinite class of 3-4-1 models can, in principle, be constructed. Furthermore, by choosing $a=1$, the Standard Model weak isospin group $SU(2)_L$ is naturally embedded into the extended gauge group $SU(4)_L$. The 15 gauge fields $A_\mu^\alpha$ of $SU(4)_L$ can be expressed as:

\setlength{\arraycolsep}{8pt}
\begin{equation}\label{eq:gaugebosons}
\frac{1}{2}\lambda_{\alpha}A_\mu^\alpha
=\frac{1}{\sqrt{2}}
\begin{pmatrix}
D_{1\mu}^{0} & W_\mu^{+} & K_\mu^{(b+1)/2}
& X_\mu^{(3+b+2c)/6}
\\[2mm]
W_\mu^{-} & D_{2\mu}^{0} & K_{1\mu}^{(b-1)/2}
& V_\mu^{(-3+b+2c)/6}
\\[2mm]
K_\mu^{-(b+1)/2} & K_{1\mu}^{-(b-1)/2} & D_{3\mu}^{0} & Y_\mu^{-(b-c)/3}
\\[2mm]
X_\mu^{-(3+b+2c)/6} & V_\mu^{(3-b-2c)/6}
& Y_\mu^{(b-c)/3} & D_{4\mu}^{0}
\end{pmatrix},
\end{equation}
\noindent here, $D_1^{0\mu} = \frac{A_3^\mu}{\sqrt{2}}
+ \frac{A_8^\mu}{\sqrt{6}}
+ \frac{A_{15}^\mu}{\sqrt{12}}$, $D_2^{0\mu} = -\frac{A_3^\mu}{\sqrt{2}}
+ \frac{A_8^\mu}{\sqrt{6}}
+ \frac{A_{15}^\mu}{\sqrt{12}}$, $D_3^{0\mu} = -\frac{2A_8^\mu}{\sqrt{6}}
+ \frac{A_{15}^\mu}{\sqrt{12}}$ and $D_4^{0\mu} = -\frac{3A_{15}^\mu}{\sqrt{12}}$. The
superscripts on the gauge bosons in Eq. \eqref{eq:gaugebosons} indicate the electric charge of the particles,
some of which are functions of the parameters b and c.

\noindent Using the explicit matrix representation of the $SU(4)_L$ Gell--Mann generators and taking $a=1$, the electric charge operator $Q$ in the fundamental representation $\mathbf{4}$ can be written as
\begin{equation}
\begin{split}
Q[\mathbf{4}] &=\frac{1}{2} \lambda_{3}
+
\frac{b}{2\sqrt{3}}\lambda_{8}+
\frac{c}{2\sqrt{6}}\lambda_{15}+
X\mathbf{1} \\
=& \mathrm{diag}\left(
\frac{1}{2}+\frac{b}{6}+\frac{c}{12}+X,\;
-\frac{1}{2}+\frac{b}{6}+\frac{c}{12}+X,\;
-\frac{b}{3}+\frac{c}{12}+X,\;
-\frac{c}{4}+X
\right)
\end{split}
\label{eq:opecarga4}
\end{equation}
\noindent
For the anti-fundamental representation $\overline{\mathbf{4}}$, the electric charge operator is diagonal and takes the form
\begin{align}
Q[\overline{\mathbf{4}}] =
\mathrm{diag}\left(
-\frac{1}{2}-\frac{b}{6}-\frac{c}{12}+X,\;
\frac{1}{2}-\frac{b}{6}-\frac{c}{12}+X,\;
\frac{b}{3}-\frac{c}{12}+X,\;
\frac{c}{4}+X
\right).\label{eq:opecarga4*}
\end{align}
\section{fermion families   for \texorpdfstring{$b=c=1$}{}}
\noindent In accordance with the definition of the electric charge operator given in Eqs.\eqref{eq:opecarga4} and \eqref{eq:opecarga4*}, the different multiplets (quartets of $SU(4)_L$) of the 3-4-1 model with $b=c=1$ are constructed by evaluating the charge operator $Q$ in the $\mathbf{4}$ and $\overline{\mathbf{4}}$ representations for different values of the $U(1)_X$ charge quantum number $X$, in such a way that the Standard Model particle content is reproduced in the first two components of the $SU(4)_L$ quartets.\\

\noindent Accordingly, the left-handed quark and lepton quartets 
transforming as the fundamental representation $\mathbf{4}$ of $SU(4)_L$ and associated with each Standard Model family must have the generic structure: 
\[
(\nu_{e},\,e^{-},\,L,\,L')_{L}^{T},
\qquad
(u,\,d,\,Q,\,Q')_{L}^{T},
\]
where $L$ and $L'$ denote exotic leptons, while $Q$ and $Q'$ denote exotic quarks. Likewise, for the conjugate representation $\overline{\mathbf{4}}$, the left-handed lepton and quark quartets take the generic form
\[
(e^{-},\,\nu_{e},\,L,\,L')_{L}^{T},
\qquad
(d,\,u,\,Q,\,Q')_{L}^{T}.
\]
Notice that the electric charges of the exotic quarks and leptons differ between the fundamental and conjugate representations.

\noindent For the particular choice $b=c=1$, the electric charge operator in the $\mathbf{4}$ and $\overline{\mathbf{4}}$ representations is given by
\begin{subequations}
\begin{align}
Q[\mathbf{4}]&=
\mathrm{Diag}\!\left(
\frac{3}{4}+X,\,
-\frac{1}{4}+X,\,
-\frac{1}{4}+X,\,
-\frac{1}{4}+X
\right),\\ 
Q[\overline{\mathbf{4}}]&=
\mathrm{Diag}\!\left(
-\frac{3}{4}+X,\,
\frac{1}{4}+X,\,
\frac{1}{4}+X,\,
\frac{1}{4}+X
\right).
\end{align}
\end{subequations}

We construct the fermion spectra from quark and lepton multiplets, with left-handed fields transforming as quadruplets of $SU(4)_L$ and right-handed fields as singlets under $SU(4)_L$, while avoiding exotic electric charges in the last two components of the quadruplets. The quark and lepton families are listed below.
\begin{align*}
S_{Q}^{1} &= \{ (u,d,D,D')_L \sim [3,4,-\tfrac{1}{12}],\
u^c_L \sim [\bar{3},1,-\tfrac{2}{3}],\
d^c_L \sim [\bar{3},1,\tfrac{1}{3}],\\
&\qquad D^c_L \sim [\bar{3},1,\tfrac{1}{3}],\
D'{}^c_L \sim [\bar{3},1,\tfrac{1}{3}] \}, \\[6pt]
S_{Q}^{2} &= \{ (d,u,U,U')_L \sim [3,\bar{4},\tfrac{5}{12}],\
u^c_L \sim [\bar{3},1,-\tfrac{2}{3}],\
d^c_L \sim [\bar{3},1,\tfrac{1}{3}],\\
&\qquad U^c_L \sim [\bar{3},1,-\tfrac{2}{3}],\
U'{}^c_L \sim [\bar{3},1,-\tfrac{2}{3}] \}, \\[6pt]
S_{L}^{3} &= \{ (\nu_e,e^-,E^-,E'^-)_L \sim [1,4,-\tfrac{3}{4}],\
e^+_L \sim [1,1,1],\
E^+_L \sim [1,1,1],\
E'^+_L \sim [1,1,1] \}, \\[6pt]
S_{L}^{4} &= \{ (E^+,N^0_1,N^0_2,N^0_3)_L \sim [1,4,\tfrac{1}{4}],\
E^-_L \sim [1,1,-1] \}, \\[6pt]
S_{L}^{5} &= \{ (e^-,\nu_e,N^0,N'^0)_L \sim [1,\bar{4},-\tfrac{1}{4}],\
e^+_L \sim [1,1,1] \}, \\[6pt]
S_{L}^{6} &= \{ (N^0,E^+_1,E^+_2,E^+_3)_L \sim [1,\bar{4},\tfrac{3}{4}],\
E^-_{1L},E^-_{2L},E^-_{3L} \sim [1,1,-1] \}.
\end{align*}
The contributions of each quark and lepton set to
the gauge anomalies are summarized in Table~\ref{tab:Anomalia_bc1}.
\begin{table}[H]
\centering
\caption{Contribution to the anomalies for each family of quarks $S_{Qi}$ and leptons $S_{Li}$ for 3-4-1 models with $b = c = 1$.}
\label{tab:Anomalia_bc1}
\setlength{\tabcolsep}{15pt}
\renewcommand{\arraystretch}{1.9}
\begin{tabular}{lcccccc}
\hline\hline
\textbf{Anomaly} 
& $S_Q^1$ & $S_Q^2$ 
& $S_L^3$ & $S_L^4$ & $S_L^5$ & $S_L^6$ \\
\hline
$[U(1)_X]^3$ 
& $-\dfrac{9}{16}$ 
& $-\dfrac{27}{16}$ 
& $\dfrac{21}{16}$ 
& $-\dfrac{15}{16}$ 
& $\dfrac{15}{16}$ 
& $-\dfrac{21}{16}$ \\
$[SU(4)_L]^2 U(1)_X$ 
& $-\dfrac{1}{4}$ 
& $\dfrac{5}{4}$ 
& $-\dfrac{3}{4}$ 
& $\dfrac{1}{4}$ 
& $-\dfrac{1}{4}$ 
& $\dfrac{3}{4}$ \\
$[SU(4)_L]^3$ 
& $3$ & $-3$ 
& $1$ & $1$ 
& $-1$ & $-1$ \\
\hline\hline
\end{tabular}
\end{table}

\noindent For the SM fields embedded in $S_L^3$, $S_L^5$, $S_Q^1$, and
$S_Q^2$, the $Z'$ charges are shown in tables \ref{tab:cargasSL3}–\ref{tab:cargasSQ2}, respectively.

\section{fermion families   for \texorpdfstring{$b=1$}{} and \texorpdfstring{$c=-2$}{}}
For $b=1$ and $c=-2$, the electric charge operator in the $\textbf{4}$ and $\overline{\textbf{4}}$ representations can be written as:
\begin{subequations}
\begin{align}
Q[\textbf{4}]=& \mathrm{diag} \biggr(\frac{1}{2}+ X, -\frac{1}{2}+X, -\frac{1}{2} +X, \frac{1}{2}+X\biggl)
\hspace{0.3cm}\\
Q[\overline{\textbf{4}}]=& \mathrm{diag} \biggr(-\frac{1}{2}+ X, \frac{1}{2}+X, \frac{1}{2} +X, -\frac{1}{2}+X\biggl)
\end{align}
\end{subequations}
From these expressions, the electric charge assignments for the fields in the fundamental and antifundamental representations can be readily obtained. For the quark multiplets, one finds
\begin{align}
Q[\textbf{4}]_{X=1/6} &=
\left(\frac{2}{3},\,-\frac{1}{3},\,-\frac{1}{3},\,\frac{2}{3}\right),\\
Q[\overline{\textbf{4}}]_{X=1/6} &=
\left(-\frac{1}{3},\,\frac{2}{3},\,\frac{2}{3},\,-\frac{1}{3}\right).
\end{align}
These correspond to quark multiplets of the form
$
(u,d,D, U)^T\quad \text{and} \quad
(d,u,U,D)^T$,
respectively. Here, ($U$) and ($D$) denote heavy exotic quarks.
Similarly, for the lepton multiplets,
\begin{align}
Q[\textbf{4}]_{X=-1/2} &=
\left(0,\,-1,\,-1,\,0\right),\\
Q[\overline{\textbf{4}}]_{X=-1/2} &=
\left(-1,\,0,\,0,\,-1\right).
\end{align}
which correspond to fermion multiplets of the form
$
(\nu^0, e^-, E^-, N^0)^T
\quad \text{and} \quad
(e^-, \nu^0, N^0, E^-)^T$,
respectively. Here, ($E^-$) and ($N^0$) denote heavy exotic charged leptons and neutral leptons, respectively.
To ensure anomaly cancellation, it is convenient to introduce exotic lepton multiplets. In particular, we consider the multiplets $(E^+,N_{1}^0,N_{2}^0,e^{+})^T$, transforming as a $\mathbf{4}$ with electric charges given by  $Q(\mathbf{4})_{X=1/2}$, and $(N_{3}^0,E^+,e^{+},N_{4}^0)^T$, transforming as 
a $\overline{\mathbf{4}}$ with electric charges given by 
$Q(\overline{\mathbf{4}})_{X=1/2}$.
\noindent We construct the fermion spectra from quark and lepton multiplets, with left-handed fields transforming as quadruplets of $SU(4)_L$ and right-handed fields as singlets under $SU(4)_L$, including exotic particles in the last two components of the quadruplets. The quark and lepton families are listed below.
\begin{align*}
S_Q^{\prime 1} &= \{ (u,d,D,U)_L \sim [3,4,\tfrac{1}{6}],\
u^c_L \sim [\bar{3},1,-\tfrac{2}{3}],\
d^c_L \sim [\bar{3},1,\tfrac{1}{3}],\\
&\qquad D^c_L \sim [\bar{3},1,\tfrac{1}{3}],\
U^c_L \sim [\bar{3},1,-\tfrac{2}{3}] \}, \\[6pt]
S_Q^{\prime 2} &= \{ (d,u,U,D)_L \sim [3,\bar{4},\tfrac{1}{6}],\
u^c_L \sim [\bar{3},1,-\tfrac{2}{3}],\
d^c_L \sim [\bar{3},1,\tfrac{1}{3}],\\
&\qquad U^c_L \sim [\bar{3},1,-\tfrac{2}{3}],\
D^c_L \sim [\bar{3},1,\tfrac{1}{3}] \}, \\[6pt]
S_L^{\prime 3} &= \{ (\nu_e,e^-,E^-,N^0)_L \sim [1,4,-\tfrac{1}{2}],\
e^+_L,E^+_L \sim [1,1,1] \}, \\[6pt]
S_L^{\prime 4} &= \{ (e^-,\nu_e,N^0,E^-)_L \sim [1,\bar{4},-\tfrac{1}{2}],\
e^+_L,E^+_L \sim [1,1,1] \}, \\[6pt]
S_L^{\prime 5} &= \{ (E^+,N^0_1,N^0_2,e^+)_L \sim [1,4,\tfrac{1}{2}],\
E^-_L,e^-_L \sim [1,1,-1] \}, \\[6pt]
S_L^{\prime 6} &= \{ (N^0_3,E^+,e^+,N^0_4)_L \sim [1,\bar{4},\tfrac{1}{2}],\
E^-_L \sim [1,1,-1],e^-_L \sim [1,1,-1] \}.
\end{align*}

\begin{table}[H]
\centering
\caption{Contribution to the anomalies for each family of quarks $S'_{Qi}$ and leptons $S'_{Li}$ for 3-4-1 models with $b = 1, c = -2$.}
\label{tab:Anomalia_b1cm2}
\setlength{\tabcolsep}{15pt}
\renewcommand{\arraystretch}{1.9}
\begin{tabular}{lcccccc}
\hline\hline
\textbf{Anomaly} 
& $S_Q'^1$ & $S_Q'^2$ 
& $S_L'^3$ & $S_L'^4$ & $S_L'^5$ & $S_L'^6$ \\
\hline
$[U(1)_X]^3$ 
& $-\dfrac{3}{2}$ 
& $-\dfrac{3}{2}$ 
& $\dfrac{3}{2}$ 
& $\dfrac{3}{2}$ 
& $-\dfrac{3}{2}$ 
& $-\dfrac{3}{2}$ \\
$[SU(4)_L]^2 U(1)_X$ 
& $\dfrac{1}{2}$ 
& $\dfrac{1}{2}$ 
& $-\dfrac{1}{2}$ 
& $-\dfrac{1}{2}$ 
& $\dfrac{1}{2}$ 
& $\dfrac{1}{2}$ \\
$[SU(4)_L]^3$ 
& $3$ & $-3$ 
& $1$ & $-1$ 
& $1$ & $-1$ \\
\hline\hline
\end{tabular}
\end{table}

\noindent For the SM lepton doublets embedded in $S_L^{\prime 3}$ and $S_L^{\prime 4}$, and the SM quark doublets embedded in $S_Q^{\prime 1}$ and $S_Q^{\prime 2}$, the corresponding $Z'$ chiral charges are given in Tables~\ref{tab:cargaSL'3} and~\ref{tab:cargaSQ'2}, respectively. The sets $S_L^{\prime 5}$ and $S_L^{\prime 6}$ contain additional leptonic states and are not represented by the SM-lepton coupling table.

\section{Anomaly-free fermion sets and minimal three-family models}
\label{sec:minimal_models}

The cancellation of gauge anomalies constrains the fermion content of
$SU(3)_C\otimes SU(4)_L\otimes U(1)_X$ theories. We construct the spectra
from the fermion sets introduced in Secs.~III and IV. For either choice
of $(b,c)$, let $S_i$ denote these six sets in the order
$(S_Q^1,S_Q^2,S_L^3,S_L^4,S_L^5,S_L^6)$, with primes understood for
$b=1$, $c=-2$. A spectrum is specified by non-negative integer
multiplicities,
\begin{equation}
  \mathcal{F}(\mathbf n)=\sum_{i=1}^{6}n_iS_i,
  \qquad \mathbf n=(n_1,\ldots,n_6)^T\in\mathbb Z_{\geq0}^{6}.
  \label{eq:fermion_multiplicities}
\end{equation}
Since the anomaly coefficients are additive, anomaly cancellation amounts to
\begin{equation}
  A\mathbf n=0,
  \label{eq:anomaly_integer_system}
\end{equation}
where the columns of $A$ are the anomaly contributions of the six sets.
The three rows correspond to $[U(1)_X]^3$,
$[SU(4)_L]^2U(1)_X$, and $[SU(4)_L]^3$; multiplying rows by
nonzero constants does not affect the solutions.

We distinguish two notions. An \emph{irreducible anomaly-free set}
(AFS) is a nonzero solution of Eq.~\eqref{eq:anomaly_integer_system}
that cannot be written as the sum of two nonzero anomaly-free spectra.
A \emph{minimal three-family model}, instead, is an anomaly-free
spectrum with at least three quark families and at least three lepton doublets
with Standard Model (SM) gauge quantum numbers, from which no nonempty
anomaly-free subset can be removed while retaining these requirements.
Additional SM-like lepton doublets are allowed as exotic fermion content;
this classification does not assume that they remain light.
In terms of multiplicities, a model $\mathbf n$ is minimal if there is
no distinct admissible solution $\mathbf m$ satisfying
$0\leq m_i\leq n_i$ for all $i$. Thus a minimal three-family model
need not be an irreducible AFS. This criterion avoids counting arbitrary
extensions obtained by adding further anomaly-free sets.

For convenience, we retain the notation $L_i$, $Q_i^{I}$,
$Q_i^{II}$, and $Q_i^{III}$ for the anomaly-free combinations with
zero, one, two, and three quark families, respectively. The primed
notation is used for the $b=1$, $c=-2$ realization.

\subsection*{Models with \texorpdfstring{$b=c=1$}{}}

Using the anomaly coefficients in Table~\ref{tab:Anomalia_bc1},
Eq.~\eqref{eq:anomaly_integer_system} becomes
\begin{equation}
 A_{(1,1)}=\begin{pmatrix}
 -9&-27&21&-15&15&-21\\
 -1&5&-3&1&-1&3\\
 3&-3&1&1&-1&-1
 \end{pmatrix}.
 \label{eq:anomaly_matrix_bc1}
\end{equation}
The anomaly-free building blocks used below are displayed in
Table~\ref{tab:AFSclasificacion1}. In particular,
$L_1=S_L^4+S_L^5$ and $L_2=S_L^3+S_L^6$ have vanishing total anomalies.

\begin{table}[H]
\centering
\caption{Irreducible anomaly-free building blocks for $b=c=1$,
classified by the number of quark families.}
\label{tab:AFSclasificacion1}
\begin{tabular}{ccccc}
\hline\hline
$i$ & Leptonic $L_i$ & One quark $Q_i^{I}$ & Two quarks $Q_i^{II}$ & Three quarks $Q_i^{III}$\\
\hline
1 & $S_L^4+S_L^5$ & $S_Q^1+2S_L^5+S_L^6$ & $S_Q^1+S_Q^2+S_L^3+S_L^5$ & $S_Q^1+2S_Q^2+3S_L^3$\\
2 & $S_L^3+S_L^6$ & $S_Q^2+2S_L^3+S_L^4$ & --- & $2S_Q^1+S_Q^2+3S_L^5$\\
\hline\hline
\end{tabular}
\end{table}

We require $n_1+n_2\geq3$ and $n_3+n_5\geq3$, since
$S_L^3$ and $S_L^5$ contain SM-like lepton doublets.
Solving Eq.~\eqref{eq:anomaly_integer_system} with these conditions
and retaining only the componentwise-minimal solutions gives the five
spectra in Table~\ref{tab:modelos2}. The first two are irreducible
AFSs. The remaining three are algebraically reducible: two consist of
three one-quark AFSs, while the fifth spectrum consists of two
two-quark AFSs and therefore contains four quark families. Removing
any nonempty anomaly-free subset from these spectra necessarily leaves
fewer than three quark families or fewer than three SM-like lepton
doublets. Thus all five are minimal in the sense defined above.

\begin{table}[H]
\centering
\caption{Minimal three-family fermion spectra for $b=c=1$.
The five minimal models are numbered consecutively.}
\label{tab:modelos2}
\begin{tabular}{ccll}
\hline\hline
Model & AFS decomposition & Fermion spectrum & Classification\\
\hline
$M_1$ & $Q_1^{III}$ & $S_Q^1+2S_Q^2+3S_L^3$ & Irreducible\\
$M_2$ & $Q_2^{III}$ & $2S_Q^1+S_Q^2+3S_L^5$ & Irreducible\\
$M_3$ & $3Q_1^{I}$ & $3S_Q^1+6S_L^5+3S_L^6$ & Reducible, minimal\\
$M_4$ & $3Q_2^{I}$ & $3S_Q^2+6S_L^3+3S_L^4$ & Reducible, minimal\\
$M_5$ & $2Q_1^{II}$ & $2S_Q^1+2S_Q^2+2S_L^3+2S_L^5$ & Reducible, minimal\\
\hline\hline
\end{tabular}
\end{table}

The reducible spectra excluded from Table~\ref{tab:modelos2}
contain proper anomaly-free subsets whose removal preserves the
required family content. In particular, adding $L_2$ or $L_1$
to $M_2$ or $M_1$ produces nonminimal extensions. More generally,
there are infinitely many such extensions. The extra quark and lepton
doublets in the minimal spectra are counted by their SM gauge quantum
numbers; whether they can acquire sufficiently large masses depends
on the scalar and Yukawa sectors and is not established by anomaly
cancellation alone.

\subsection*{Models with \texorpdfstring{$b=1$, $c=-2$}{}}

For this realization, the anomaly coefficients of
Table~\ref{tab:Anomalia_b1cm2} give
\begin{equation}
 A_{(1,-2)}=\begin{pmatrix}
 -3&-3&3&3&-3&-3\\
 1&1&-1&-1&1&1\\
 3&-3&1&-1&1&-1
 \end{pmatrix},
 \label{eq:anomaly_matrix_b1cm2}
\end{equation}
where the first row has been multiplied by two and the second by two.
The corresponding anomaly-free building blocks are given in
Table~\ref{tab:AFSclasificacion2}. In particular,
$L'_1=S_L^{\prime4}+S_L^{\prime5}$ and
$L'_2=S_L^{\prime3}+S_L^{\prime6}$ are anomaly-free.

\begin{table}[H]
\centering
\caption{Irreducible anomaly-free building blocks for $b=1$, $c=-2$.}
\label{tab:AFSclasificacion2}
\begin{tabular}{ccccc}
\hline\hline
$i$ & Leptonic $L'_i$ & One quark $Q_i^{\prime I}$ & Two quarks $Q_i^{\prime II}$ & Three quarks $Q_i^{\prime III}$\\
\hline
1 & $S_L^{\prime4}+S_L^{\prime5}$ & $S_Q^{\prime1}+2S_L^{\prime4}+S_L^{\prime6}$ & $S_Q^{\prime1}+S_Q^{\prime2}+S_L^{\prime3}+S_L^{\prime4}$ & $S_Q^{\prime1}+2S_Q^{\prime2}+3S_L^{\prime3}$\\
2 & $S_L^{\prime3}+S_L^{\prime6}$ & $S_Q^{\prime2}+2S_L^{\prime3}+S_L^{\prime5}$ & --- & $2S_Q^{\prime1}+S_Q^{\prime2}+3S_L^{\prime4}$\\
\hline\hline
\end{tabular}
\end{table}

We now impose $n_1+n_2\geq3$ and $n_3+n_4\geq3$, since
$S_L^{\prime3}$ and $S_L^{\prime4}$ contain SM-like lepton doublets.
The five minimal solutions are listed in Table~\ref{tab:modelos}.
As before, two are irreducible AFSs and three are reducible but minimal
with respect to the three-family requirement.

\begin{table}[H]
\centering
\caption{Minimal three-family fermion spectra for $b=1$, $c=-2$.
The five minimal models are numbered consecutively.}
\label{tab:modelos}
\begin{tabular}{ccll}
\hline\hline
Model & AFS decomposition & Fermion spectrum & Classification\\
\hline
$M'_1$ & $Q_1^{\prime III}$ & $S_Q^{\prime1}+2S_Q^{\prime2}+3S_L^{\prime3}$ & Irreducible\\
$M'_2$ & $Q_2^{\prime III}$ & $2S_Q^{\prime1}+S_Q^{\prime2}+3S_L^{\prime4}$ & Irreducible\\
$M'_3$ & $3Q_1^{\prime I}$ & $3S_Q^{\prime1}+6S_L^{\prime4}+3S_L^{\prime6}$ & Reducible, minimal\\
$M'_4$ & $3Q_2^{\prime I}$ & $3S_Q^{\prime2}+6S_L^{\prime3}+3S_L^{\prime5}$ & Reducible, minimal\\
$M'_5$ & $2Q_1^{\prime II}$ & $2S_Q^{\prime1}+2S_Q^{\prime2}+2S_L^{\prime3}+2S_L^{\prime4}$ & Reducible, minimal\\
\hline\hline
\end{tabular}
\end{table}

Among the nonminimal extensions, two are obtained by adding the
anomaly-free leptonic combinations $L'_2$ and $L'_1$ to $M'_2$
and $M'_1$, respectively. Other extensions can likewise be reduced
while retaining the required three-family content. They are therefore omitted from the minimal
classification. The distinct minimal spectra may contain different
exotic fermions and need not be physically equivalent. Nevertheless,
as discussed in Appendix~\ref{sec:appendixA}, the SM fermions in this
configuration have the same relevant tree-level $Z'$ chiral charges
for the embeddings considered here. Consequently, the collider
constraints obtained from SM dilepton final states can be discussed
using a single low-energy coupling configuration, under the
assumptions specified in Sec.~VII.

\section{ \texorpdfstring{$Z^{\prime}$}{} charges for a general 3-4-1 model}

\noindent We begin with the neutral-current Lagrangian associated with the gauge group $G=SU(3)_C\otimes SU(4)_L\otimes U(1)_X$.  
The neutral current Lagrangian for these models is:
\begin{align}\label{eq:LRLagrangian}
-\mathcal{L}_{NC}
=&g_L J_{3L\mu}A_{3L}^{\mu}+g_{L}J_{8L\mu}A_{8}^{\mu}+g_{L}J_{15L\mu}A_{15}^{\mu}
+g_{X}J_{X\mu}A_{X}^{\mu}\notag \\
=&g_L J_{3L\mu}A_{3L}^{\mu}+g_{Y}J_{Y\mu} B^\mu+g_{Z'}J_{Z'\mu} Z^{\prime\mu}
 +g_{Z''}J_{Z''\mu} Z^{\prime\prime\mu}\ .
\end{align}
By means of an orthogonal matrix we can rotate the neutral gauge-boson basis $(A_{3L},A_{8L},A_{15L},A_{X})$ 
into the basis $(A_{3L},B,Z',Z'')$ where $B$ is the gauge boson associated with the hypercharge generator $Y$, i.e.,
\begin{equation}\label{rotation1}
\begin{pmatrix}
A^\mu_{3L} \\
A^\mu_{8L} \\
A^\mu_{15L}\\ 
A^\mu_{X} \\
\end{pmatrix} =
\mathcal{O} 
\begin{pmatrix}
A^\mu_{3L} \\
B^\mu \\
Z^{\prime\mu} \\
Z^{\prime\prime\mu} 
\end{pmatrix}
=
\begin{pmatrix}
1 &0 &0 &0 \\
0 &O_{11}  &O_{12}  &O_{13}  \\
0 &O_{21}  &O_{22}  &O_{23}  \\
0 &O_{31}  &O_{32}  &O_{33}  
\end{pmatrix}
\begin{pmatrix}
A^\mu_{3L} \\
B^\mu \\
Z^{\prime\mu} \\
Z^{\prime\prime\mu} 
\end{pmatrix}\ ,
\end{equation} 
In order to keep invariant the Lagrangian, the currents must transform with the same orthogonal matrix 
\begin{align}
g_{Y}J^\mu_Y                
&= g_{L}J^\mu_{8L}O_{11} + g_{L} J^\mu_{15L} O_{21} + g_X J^\mu_X O_{31}\ , \label{Eq1a}\\ 
g_{Z^{\prime}} J_{Z^{\prime}}^\mu    
&= g_{L} J^\mu_{8L}O_{12} + g_{L} J^\mu_{15L} O_{22} + g_X J^\mu_X O_{32}\ , \label{Eq1b}\\
g_{Z^{\prime\prime}}J_{Z^{\prime\prime}}^\mu   
&= g_{L} J^\mu_{8L}O_{13} + g_{L} J^\mu_{15L} O_{23} + g_X J^\mu_X O_{33}\ . \label{Eq1c} 
\end{align}
From the general charge operator, we have 
\begin{align}
Y= \beta T_{8L} + \gamma T_{15L}+X\mathbf{1}\ .    
\end{align}
This implies the relation
\begin{align}
J_Y= \beta J_{8L} + \gamma J_{15L}+J_X\ . 
\label{eq:jy}
\end{align}
 From comparison with the charge operator \eqref{qem}, and using the relation $Q=T_{3L}+Y$, the following relations follow:
\begin{equation}\label{qem2}
\beta= \frac{b}{\sqrt{3}} \hspace{1.0cm},\hspace{1.0cm} \gamma= \frac{c}{\sqrt{6}}\ .
\end{equation}
By comparing $J_Y$ in~\eqref{Eq1a} with Eq.~\eqref{eq:jy} we  obtain 
\begin{align}\label{eq:angles1}
\frac{g_{L}}{g_Y}O_{11}= \beta\ ,
\hspace{0.5cm}
\frac{g_L}{g_Y}O_{21}= \gamma\ ,
\hspace{0.5cm}
\frac{g_{X}}{g_Y}O_{31}= 1\ .
\end{align}
As is well known, an orthogonal matrix satisfies
$(O^{T}O)_{ij}=O_{ki}O_{kj}=\delta_{ij}$. For $i=j=1$ we have   
$O_{k1}O_{k1}=O_{11}^2+O_{21}^2+O_{31}^2 =1$, 
such that 
\begin{align}\label{ortogonal}
\left(\frac{\beta}{g_{L}}\right)^2
+\left(\frac{\gamma}{g_L}\right)^2
+\left(\frac{1}{g_X}\right)^2=\frac{1}{g_Y^2}\ . 
\end{align}
Let us now consider an explicit representation for the orthogonal matrix $\mathcal{O}$ in terms 
of three angles $\omega$, $\phi$ and $\psi$, which are allowed to take values in the $[-\pi,\pi)$ interval. For convenience we choose

\begin{align}
\mathcal{O} &=
\begin{pmatrix}
1 &0  &0  &0 \\
0 &\cos w &-\sin w &0 \\
0 &\sin w &\cos w &0 \\
0 &0 &0 &1
\end{pmatrix}
\begin{pmatrix}
1 &0  &0  &0 \\
0 &\cos\phi &0 &-\sin\phi \\
0 &0 &1 &0 \\
0 &\sin\phi &0 &\cos\phi
\end{pmatrix}
\begin{pmatrix}
1 &0  &0  &0 \\
0 &1 &0 &0 \\
0 &0 &\cos\psi &-\sin\psi \\
0 &0 &\sin\psi &\cos\psi
\end{pmatrix}\ .\notag 
\end{align}

From Eq.~\eqref{eq:angles1} we obtain 
\begin{align}\label{identidades}
\cos\phi \cos\omega= \frac{\beta g_Y}{g_{L}}\ ,
\hspace{0.3cm}
\cos\phi \sin\omega= \frac{\gamma g_Y}{g_{L}}\ , 
\hspace{0.3cm}
\sin\phi = \frac{ g_Y}{g_{X}}\ , 
\end{align}


By assuming 
\begin{align}
J_{Z'}^{\mu}=&\sum_f \bar{f}\gamma^\mu\left(\epsilon^{Z'}_L(f)P_L+\epsilon^{Z'}_R(f)P_R\right)f\ ,\notag\\
J_{8L}^{\mu}=&\sum_f \bar{f}\gamma^\mu\epsilon^{8}_{L}(f)P_Lf\ ,\notag\\
J_{X}^{\mu}=&\sum_f \bar{f}\gamma^\mu\left(X_{L}(f)P_L+X_R(f)P_R\right)f\ ,
\end{align}
where we define $\epsilon_{L,R}^a$ as the corresponding 
chiral charges associated with the gauge group generators $T_{L,R}^{a}=\frac{1}{2}\lambda^{a}$, with $\lambda^{a}$ 
the Gell-Mann matrices. 
Replacing  these expressions in Eq.~\eqref{Eq1b} and using  the elements of the rotation matrix $\mathcal{O}$  we get
the $Z'$ chiral charges:
\begin{align}
  g_{Z'} \epsilon_{L}^{Z'}&= A_{L}\cos{\psi}+ B_{L}\sin{\psi} \label{eq:zizq} \\
  g_{Z'} \epsilon_{R}^{Z'}&= A_{R}\cos{\psi}+ B_{R}\sin{\psi} \label{eq:zder}\ .
\end{align}
\noindent
Here the angle \(\psi\) parameterizes the rotation between the additional neutral gauge-boson directions orthogonal to hypercharge and should not be confused with the mixing angle between the Standard Model \(Z\) boson and the additional \(Z'\), which is neglected in our collider analysis.
In Eqs.~\eqref{eq:zizq} and~\eqref{eq:zder}, the functions 
 $A_{L,R}$ and $B_{L,R}$ are defined as:
\begin{subequations}
    \begin{align}
       A_{L} & = \frac{1}{\tilde \alpha_{L}}\left(\beta \epsilon_{L}^{15}- \gamma\epsilon_{L}^{8}  \right)\ , \\
       B_{L} & = g_{Y}\left(g_{X}X_{L}\tilde \alpha_{L}- \frac{\beta}{g_{X}\tilde \alpha_{L}} \epsilon_{L}^{8}- \frac{\gamma}{g_{X}\tilde \alpha_{L}} \epsilon_{L}^{15}  \right)\ ,\\ 
       A_{R} & = 0\ , \\
       B_{R} & = g_{Y}g_{X}X_{R}\tilde \alpha_{L}\ .
    \end{align}
\end{subequations}
The quantities $\tilde\alpha_{L}$ and $g_{X}$ are determined from Eqs.~\eqref{ortogonal} and~\eqref{identidades} as follows: 
\begin{align}
    \tilde\alpha_{L} &=\frac{1}{g_{L}}\sqrt{\beta ^2 + \gamma ^2}\label{eq:alfa}\ ,\\
    g_{X} &= \frac{g_{L} \tan{\theta_{W}}}{\sqrt{1-(\beta ^2 + \gamma ^2)\tan^{2}{\theta_{W}}}}\label{eq:gx}\ .
\end{align}
\noindent Here $X_R$ denotes the $U(1)_X$ charge of the right-handed $SU(4)_L$ singlets. Its normalization, and consequently that of the gauge coupling $g_X$, depends on the definition of the electric charge operator given in Eq.~\eqref{qem}.\\

Since $\epsilon^{i}=T^{i}=\frac{1}{2}\lambda^{i}$ are the generators of $SU(4)_L$, Eq.~\eqref{eq:zizq} can be expressed in matrix form as follows:
\setlength{\arraycolsep}{2pt}
\begin{align}
  &g_{Z'} \epsilon_{L}^{Z'}= \frac{1}{\tilde \alpha_{L}}\left(\beta \epsilon_{L}^{15}- \gamma\epsilon_{L}^{8}  \right)\cos{\psi}+g_{Y} \left(g_{X}X_{L}\tilde \alpha_{L}- \frac{\beta}{g_{X}\tilde \alpha_{L}} \epsilon_{L}^{8}- \frac{\gamma}{g_{X}\tilde \alpha_{L}} \epsilon_{L}^{15}  \right)\sin{\psi}\notag \\
 &=\frac{1}{\tilde \alpha_{L}}
\left(
\begin{array}{cccc}
 \frac{\beta }{2 \sqrt{6}}-\frac{\gamma }{2 \sqrt{3}} & 0 & 0 & 0 \\
 0 & \frac{\beta }{2 \sqrt{6}}-\frac{\gamma }{2 \sqrt{3}} & 0 & 0 \\
 0 & 0 & \frac{\beta }{2 \sqrt{6}}+\frac{\gamma }{\sqrt{3}} & 0 \\
 0 & 0 & 0 & -\frac{1}{2} \sqrt{\frac{3}{2}} \beta  \\
\end{array}
\right)\cos{\psi}+\tilde \alpha_{L}g_{Y} g_{X} X_{L}\left(
\begin{array}{cccc}
  1 & 0 & 0 & 0  \\
  0 & 1 & 0 & 0  \\
  0 & 0 & 1 & 0  \\
  0 & 0 & 0 & 1  \\
\end{array}
\right)\sin{\psi} \notag\\
&+ g_{Y}
{\setlength{\arraycolsep}{0pt}
\left(
\begin{array}{cccc}
 -\frac{\beta }{2 \sqrt{3} \tilde \alpha_{L} g_{X}}-\frac{\gamma }{2 \sqrt{6} \tilde \alpha_{L} g_{X}} & 0 & 0 &
   0 \\
 0 & -\frac{\beta }{2 \sqrt{3} \tilde \alpha_{L} g_{X}}-\frac{\gamma }{2 \sqrt{6}\tilde \alpha_{L} g_{X}} & 0 &
   0 \\
 0 & 0 & \frac{\beta }{\sqrt{3} \tilde \alpha_{L} g_{X}}-\frac{\gamma }{2 \sqrt{6} \tilde \alpha_{L} g_{X}} & 0 \\
 0 & 0 & 0 & \frac{\sqrt{\frac{3}{2}} \gamma }{2 \tilde \alpha_{L} g_{X}} \\
\end{array}
\right)}\sin{\psi}\ .
 \end{align}

\noindent From Eq.\eqref{eq:zder} the right-handed chiral charge of the $Z'$ is written as:
\begin{equation}
g_{Z'} \epsilon_{R}^{Z'}= g_{Y}g_{X}X_{R}\tilde \alpha_{L}
\sin{\psi}.
\end{equation}

\section{Collider Constraints}
\label{sec:collider}

To investigate the collider phenomenology of the
$SU(3)_C\otimes SU(4)_L\otimes U(1)_X$ models,
we consider the ATLAS search for heavy neutral
resonances decaying into opposite-sign electron
and muon pairs~\cite{ATLAS:2019erb}.
This analysis uses the full Run~2 dataset of
proton--proton collisions at $\sqrt{s}=13$ TeV,
corresponding to an integrated luminosity of
$139\,\mathrm{fb}^{-1}$, and covers dilepton
invariant masses between $250$ GeV and $6$ TeV.
No statistically significant excess above the
Standard Model (SM) expectation was observed,
allowing the collaboration to establish upper
limits at the $95\%$ confidence level on the
production cross section times the dilepton
branching ratio.

In deriving the collider constraints, we adopt
several simplifying assumptions. First, the
right-handed neutrinos and all exotic fermions
are assumed to be sufficiently heavy that their
decay channels are kinematically inaccessible
to the $Z'$ boson. Formally, this corresponds
to taking their masses to infinity. Consequently,
only decays into SM fermions are included in the
width used in our theoretical dilepton cross sections.
Possible decays into scalar or gauge bosons are not
included; our numerical limits therefore assume that
such nonfermionic contributions to the total width are
negligible.  Second, we neglect the mixing between the SM
$Z$ boson and the additional neutral gauge boson
$Z'$~\cite{Erler:2009jh}. This is a standard approximation in
collider analyses of additional neutral gauge
bosons and allows us to investigate the direct
constraints associated with the $Z'$ fermionic
couplings without introducing additional
electroweak mixing effects.

Finally, to extract the experimental constraints,
we use the ATLAS upper limits corresponding to
a narrow neutral resonance with a width of
approximately $1\%$ of its mass,
$\frac{\Gamma_{Z'}}{M_{Z'}}\sim 0.01$,
which is representative of the narrow resonances
commonly encountered in $E_6$-inspired models.
This width is used to select the experimental
signal hypothesis and should be distinguished
from the theoretical width predicted for each
specific configuration of the $3$-$4$-$1$ model.

Although the fermionic width $\Gamma_{Z'}$ used in our calculation does not necessarily coincide with the benchmark width assumed in the ATLAS analysis, the experimental upper limits on the fiducial cross section exhibit only a weak dependence on $\Gamma_{Z'}$ in the narrow-width regime, as shown in Fig.~3 of Ref.~\cite{ATLAS:2019erb}. Consequently, provided that the predicted resonance remains sufficiently narrow, the ATLAS upper limits can be reliably applied to our model, even though the theoretical cross section depends strongly on $\Gamma_{Z'}$.

For each value of the mixing angle $\psi$, we
calculate the theoretical dilepton production
cross section~\cite{Erler:2011ud,Rojas:2015tqa,Salazar:2015gxa,Benavides:2018fzm,Suarez:2023ozu,Benavides:2025idi}.
The lower bound on the $Z'$ mass is then
determined from the intersection of the
theoretical prediction with the experimental
$95\%$ confidence level upper limit.
The resulting bounds as functions of $\psi$
are presented in Fig.~\ref{fig:colliders}.

An important result of our analysis is that
the collider phenomenology depends strongly
on the parameters $b$ and $c$ defining the
electric charge operator. In particular,
the two configurations considered in this
work exhibit qualitatively different behavior.

\subsection{The \texorpdfstring{$b=c=1$}{} configuration:
a fermiophobic limit}

For $b=c=1$, corresponding to
$
\beta=\frac{1}{\sqrt{3}}$,$\gamma=\frac{1}{\sqrt{6}},
$
the chiral couplings of the $Z'$ boson to
all SM fermions are proportional to
$\sin\psi$, as shown explicitly in
Appendix~\ref{sec:appendixA}.
The corresponding neutral-current interaction
can therefore be expressed as
\begin{equation}
\mathcal{L}_{Z'}^{\mathrm{SM}}
=
-Z'_\mu
\sum_{f\in\mathrm{SM}}
\bar f\gamma^\mu
\left(
 g_{Z'}\epsilon_L^{Z'}(f) P_L+
 g_{Z'}\epsilon_R^{Z'}(f) P_R
\right)f,
\label{eq:fermiophobic_current}
\end{equation}
where $ g_{Z'}\epsilon^{Z'}_{L,R}(f)\propto \sin\psi$.
Consequently, all the corresponding couplings
vanish simultaneously when
$
\psi=n\pi,\text{with}\  n\in\mathbb{Z}.
$
At these particular values of the mixing
angle, the $Z'$ boson becomes fermiophobic
with respect to the SM fermions. Remarkably,
this property is independent of the specific
low-energy fermion embedding considered
within the $b=c=1$ class.
The phenomenological consequences are
particularly interesting. The conventional
Drell--Yan production mechanism,
$
q\bar q\longrightarrow Z'
\longrightarrow\ell^+\ell^-,
$
requires nonvanishing couplings to both
quarks and charged leptons. Therefore,
the dilepton signal is strongly suppressed
as the mixing angle approaches the
fermiophobic values.
Under our assumption that only SM fermionic
decay channels are accessible, the partial
widths satisfy
$
\Gamma(Z'\rightarrow f\bar f)
\propto\sin^2\psi.
$
Consequently, the total fermionic width
obeys the same scaling,
$
\Gamma_{Z'}\propto\sin^2\psi.
$
Within the narrow-width approximation,
the dilepton production cross section
therefore scales as
$\sigma(pp\rightarrow Z'\rightarrow
\ell^+\ell^-)
\propto\sin^2\psi,
$
provided that the branching ratios remain
finite in the limit of vanishing couplings.
This behavior is directly reflected in
Fig.~\ref{fig:colliders}, where the lower
bounds on $M_{Z'}$ decrease significantly
as $\psi$ approaches the fermiophobic
values. At the exact decoupling points,
the conventional dilepton channel becomes
insensitive to the $Z'$ boson.
This constitutes one of the central results
of our analysis. Although additional neutral
gauge bosons arising from extended gauge
symmetries generally possess appreciable
couplings to SM fermions, the $b=c=1$
configuration provides an explicit
counterexample. In this case, a neutral
gauge boson associated with an extended
non-Abelian electroweak symmetry can
become effectively invisible to conventional
fermionic resonance searches.
The implications are particularly relevant
for the possible mass scale of the additional
neutral gauge boson. While collider searches
typically constrain $Z'$ bosons with
electroweak-size couplings to lie at the
multi-TeV scale, the fermiophobic limit
allows these direct dilepton constraints
to be evaded. Therefore, considerably
lighter $Z'$ bosons cannot be excluded
on the basis of these searches alone.

The general class of $3$-$4$-$1$ models
without exotic electric charges has been
previously investigated in
Refs.~\cite{Ponce:2003uu,Sanchez:2004uf,Ponce:2006vw,Sanchez:2008qv,Nisperuza:2009xm,Villada:2009iu}.
In particular, Ref.~\cite{Nisperuza:2009xm}
provides an explicit example in the $b=c=1$ class
in which the vacuum alignment $v=v'$ and $V=V'$
causes an additional neutral gauge boson, denoted
$Z''$ in that reference, to decouple from the other
neutral states and to couple only to exotic fermions.
This demonstrates that a neutral gauge-boson
direction without couplings to ordinary SM fermions
can arise from a concrete symmetry-breaking sector.
The identification of this direction with the $Z'$
parameterized by $\psi$ in our analysis, however,
depends on the neutral gauge-boson mass matrix and
the chosen scalar sector. Our collider limits treat
$\psi$ as a phenomenological parameter; determining
which values are realized for a given scalar sector,
and whether a light fermiophobic state satisfies all
other constraints, is left for future work.

\subsection{The \texorpdfstring{$b=1$, $c=-2$}{} configuration}

A qualitatively different situation arises
for the configuration
$
b=1,\qquad c=-2$,
or equivalently,
$
\beta=\frac{1}{\sqrt{3}}$, $
\gamma=-\frac{2}{\sqrt{6}}.
$
In this case, the chiral couplings of the
$Z'$ boson contain contributions proportional
to both $\cos\psi$ and $\sin\psi$, as follows
from Eqs.~\eqref{eq:zizq} and
\eqref{eq:zder}:
\begin{equation}
g_{Z'}\epsilon_{L,R}^{Z'}(f)
=
A_{L,R}^{f}\cos\psi+
B_{L,R}^{f}\sin\psi.
\label{eq:couplings_b1cm2}
\end{equation}

Unlike the previous configuration, there
is no common value of $\psi$ for which
all the couplings to SM fermions vanish
simultaneously. Thus, the $b=1$, $c=-2$
configuration does not exhibit the
universal fermiophobic limit found
for $b=c=1$.
As a consequence, the $Z'$ boson retains
nonvanishing couplings to SM fermions
throughout the mixing-angle parameter
space, and conventional Drell--Yan
dilepton searches remain sensitive to
its presence. This corresponds to the
more familiar collider phenomenology
of additional neutral gauge bosons
arising from extended gauge symmetries.

An additional important result concerns
the classification of anomaly-free
fermion spectra. As discussed in the
previous sections, several distinct
anomaly-free models can be constructed
for $b=1$, $c=-2$. Nevertheless, the
different embeddings of the SM fermions
lead to the same relevant low-energy
$Z'$ charge configuration, as shown
in Appendix~\ref{sec:appendixA}.
Consequently, although several anomaly-free
models exist at the level of the complete
fermionic spectrum, only one independent
configuration of low-energy $Z'$ couplings
needs to be considered for the collider
analysis.
This equivalence is particularly useful
under the assumptions adopted here.
Since all exotic fermions are taken to
be kinematically inaccessible, their
different representations and multiplicities
do not contribute to the $Z'$ decay width.
The collider phenomenology is therefore
determined entirely by the common
low-energy couplings to SM fermions.
Hence, the collider constraints obtained
for this configuration apply to all the
corresponding anomaly-free models within
the approximations of our analysis.

\subsection{Phenomenological implications\label{sec:pheno}}

The comparison between the two configurations
reveals a remarkable feature of the general
$3$-$4$-$1$ framework. Models based on the
same extended gauge symmetry can exhibit
qualitatively different collider phenomenology,
depending on the embedding of the electromagnetic
charge operator and the neutral gauge-boson
mixing structure.
For $b=c=1$, the existence of a universal
fermiophobic limit provides a mechanism
through which conventional dilepton
resonance bounds can be evaded. In contrast,
for $b=1$, $c=-2$, the absence of such a
limit leads to the more conventional
behavior expected for additional neutral
gauge bosons.
These results demonstrate that collider
bounds on the mass of a $Z'$ boson cannot
be inferred solely from the underlying
gauge group. The fermion embeddings,
the neutral gauge-boson mixing angle,
and the resulting chiral couplings
play an essential role.
In particular, the fermiophobic
configuration opens an interesting
possibility for comparatively light
neutral gauge bosons within extended
electroweak gauge theories. Establishing
the viability of such scenarios requires
a dedicated analysis of the scalar
sector and the complete set of
phenomenological constraints.

\begin{widetext}
\begin{center}
 \includegraphics[scale=0.4]{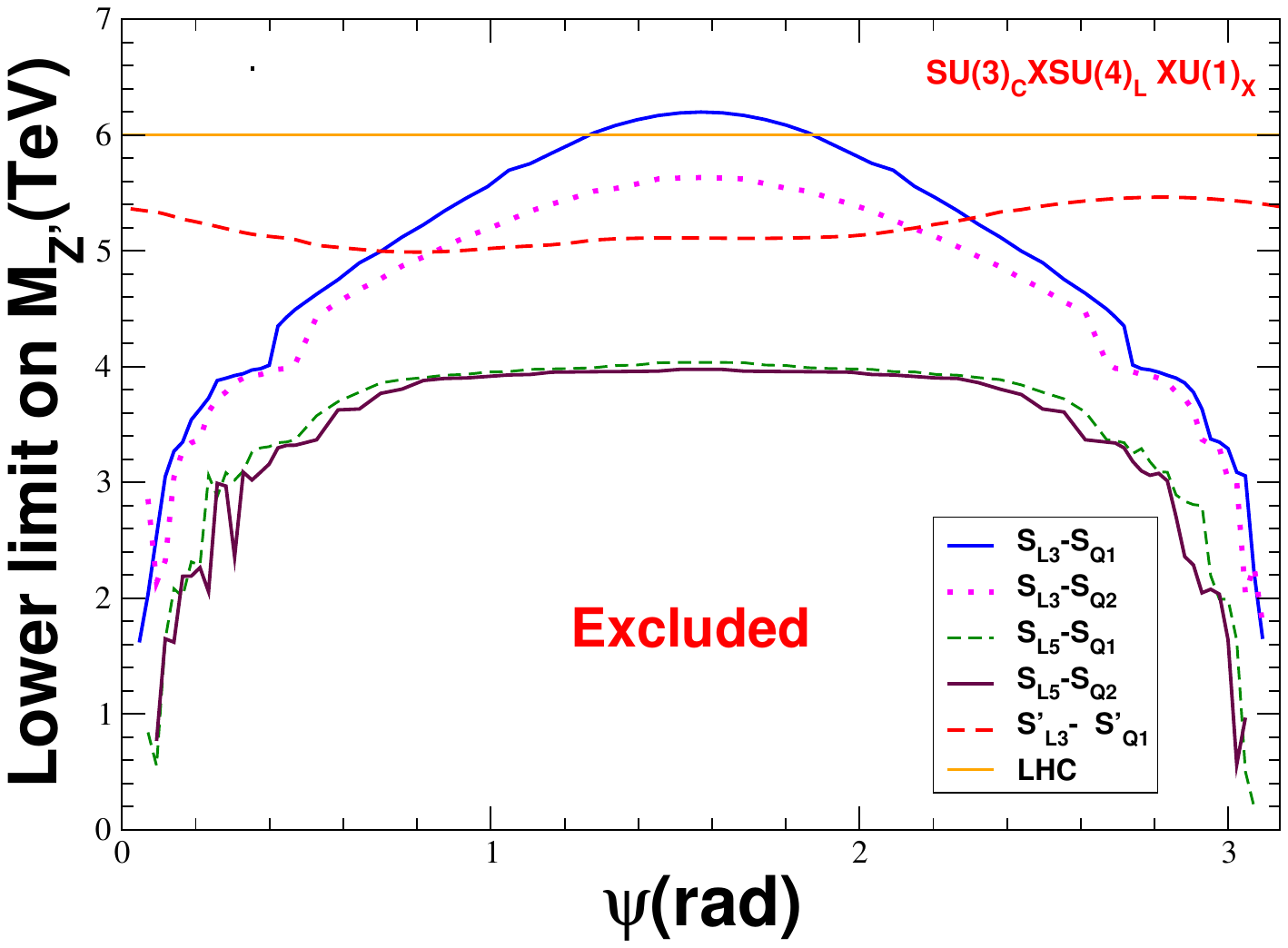}  
\end{center}
 \captionof{figure}{
Lower bounds on the $Z'$ mass as a function
of the mixing angle $\psi$, obtained from
the ATLAS $95\%$ CL upper limits on
dilepton resonance production
\cite{ATLAS:2019erb}.
The bounds are determined by the intersections
of the theoretical predictions for
$\sigma(pp\to Z'\to\ell^+\ell^-)$
with the experimental upper-limit curve,
using the narrow-resonance hypothesis
with $\Gamma_{Z'}/M_{Z'}\sim 1\%$.
The different curves correspond to the
low-energy fermion charge configurations
identified by their lepton and quark
multiplets. Primed sets denote the
$b=1$, $c=-2$ configuration.
The corresponding chiral charges are
given in Appendix~\ref{sec:appendixA}.
The fermiophobic behavior of the
$b=c=1$ configurations is reflected
in the weakening of the mass limits
near $\psi=n\pi$. The horizontal line labeled
``LHC'' indicates the upper end (approximately
$6$~TeV) of the $Z'$ mass range covered by the
ATLAS search used here; it is not an independent
exclusion limit. For each model curve, masses below
the corresponding lower-bound curve are excluded
within the assumptions of this analysis.
}
\label{fig:colliders}
\end{widetext}

\section{Conclusions}
\label{sec:conclusions}

In this work, we have performed a systematic
classification of anomaly-free fermion spectra
in models based on the
$SU(3)_C\otimes SU(4)_L\otimes U(1)_X$
gauge symmetry. We have considered two
configurations of the electric charge operator,
$b=c=1$ and $b=1$, $c=-2$, and identified
the corresponding irreducible anomaly-free
fermion sets. Imposing the requirement of at least
three quark families and three SM-like lepton
doublets, we identify five minimal models for
each configuration: two algebraically irreducible
and three algebraically reducible models that
remain minimal under the imposed family
requirements.

Starting from the general neutral-current
Lagrangian, we have derived the chiral
couplings of the additional neutral gauge
boson $Z'$ to Standard Model (SM) fermions.
An important result is that the number of
independent low-energy charge configurations
is smaller than the number of anomaly-free
models. In particular, for $b=1$, $c=-2$,
the different fermion embeddings lead to
the same relevant $Z'$ couplings to SM
fermions. Consequently, only one
independent low-energy charge configuration
needs to be considered in the collider
analysis of this class of models.

We have investigated the collider constraints
using the ATLAS dilepton resonance search
at $\sqrt{s}=13$ TeV, corresponding to an
integrated luminosity of
$139\,\mathrm{fb}^{-1}$.
The analysis assumes vanishing $Z-Z'$
mixing and sufficiently heavy right-handed
neutrinos and exotic fermions, such that
their decay channels are kinematically
inaccessible. The lower bounds on the
$Z'$ mass were obtained from the
intersection of the theoretical dilepton
production cross sections with the
experimental $95\%$ confidence level
upper limits, adopting the approximately
$1\%$ resonance-width hypothesis.

The most remarkable result of our analysis
is the existence of a fermiophobic limit
for the $b=c=1$ configuration. In this
case, all the $Z'$ couplings to SM
fermions are proportional to $\sin\psi$
and therefore vanish simultaneously
at $\psi=n\pi$, where $n$ is an integer.
As a consequence, the conventional
Drell--Yan dilepton signal disappears
at these values of the mixing angle,
and the corresponding collider bounds
on the $Z'$ mass cease to apply.

This result is particularly interesting
because it demonstrates that an additional
neutral gauge boson arising from an
extended non-Abelian electroweak gauge
symmetry does not necessarily possess
appreciable couplings to SM fermions.
Therefore, the absence of a dilepton
resonance cannot, by itself, exclude
the existence of a comparatively light
$Z'$ boson within this framework.

In contrast, for $b=1$, $c=-2$,
the chiral couplings contain contributions
proportional to both $\cos\psi$ and
$\sin\psi$, and no common value of
$\psi$ simultaneously eliminates
all the couplings to SM fermions.
Consequently, this configuration
does not exhibit a universal
fermiophobic limit and retains
the more conventional collider
phenomenology of additional neutral
gauge bosons.

Our results highlight the importance
of the fermion embeddings and the
neutral gauge-boson mixing structure
in determining the collider
phenomenology of extended electroweak
gauge theories. In particular, the
fermiophobic limit identified for
$b=c=1$ provides an interesting
mechanism through which conventional
dilepton constraints can be evaded,
potentially allowing additional
neutral gauge bosons at mass scales
considerably below the usual
multi-TeV bounds.

Whether such a scenario can be realized
in a complete and phenomenologically
viable model depends on the scalar
sector and the resulting neutral
gauge-boson mass matrix. Establishing
the symmetry-breaking conditions
that lead to the fermiophobic limit,
while satisfying electroweak precision
measurements and other experimental
constraints, constitutes a natural
and particularly interesting
direction for future work.

\section*{Acknowledgments} 
We would like to thank Professors William A. Ponce and Richard Benavides for their valuable contributions to previous collaborative work on related topics. This research was partially supported by the Vicerrectoría de Investigaciones e Interacción Social (VIIS) of the Universidad de Nariño, under project numbers 3130, 3595, and 3899.

\appendix 
\section{Chiral \texorpdfstring{$Z'$}{} charges\label{sec:appendixA} }
\subsection{Configuration \texorpdfstring{$b=c=1$}{}}

For the configuration $b=c=1$, Eq.~\eqref{qem2} gives $\beta=\frac{1}{\sqrt{3}}$ and $\gamma=\frac{1}{\sqrt{6}}$. Using $\tilde\alpha_{L}=\frac{1}{\sqrt{2}g_{L}}$ from Eq.~\eqref{eq:alfa}, Eq.~\eqref{eq:zizq} becomes:
\begin{align}
 & g_{Z'} \epsilon_{L}^{Z'}={g_{L}}
\left(
\begin{array}{cccc}
0 & 0 & 0 & 0 \\
 0 & 0 & 0 & 0 \\
 0 & 0 & \frac{1}{2} & 0 \\
 0 & 0 & 0 & -\frac{1}{2}  \\
\end{array}
\right)\cos{\psi}\notag\\
&+g_{Y}
{\setlength{\arraycolsep}{1pt} 
\left(
\begin{array}{cccc}
 -\frac{1}{2 \sqrt{2}}\frac{g_{L}}{g_{X}}+\frac{1}{\sqrt{2}}\frac{g_{X}}{g_{L}}X_{L} & 0 & 0 &
   0 \\
 0 & -\frac{1}{2 \sqrt{2}}\frac{g_{L}}{g_{X}}+\frac{1}{\sqrt{2}}\frac{g_{X}}{g_{L}}X_{L} & 0 &
   0 \\
 0 & 0 & \frac{1}{2 \sqrt{2}}\frac{g_{L}}{g_{X}}+\frac{1}{\sqrt{2}}\frac{g_{X}}{g_{L}}X_{L} & 0 \\
 0 & 0 & 0 & \frac{1}{2 \sqrt{2}}\frac{g_{L}}{g_{X}}+\frac{1}{\sqrt{2}}\frac{g_{X}}{g_{L}}X_{L}
\end{array}
\right)}\sin{\psi}
\end{align}

\noindent From Eq. \eqref{eq:zder}, the right chiral  charge of $Z'$ is written as:
\begin{align}
g_{Z'} \epsilon_{R}^{Z'}= \frac{1}{\sqrt{2}}g_{X}X_{R}\tan{\theta_{W}}\sin{\psi},
\end{align}
\noindent here, $g_{Y}=g_{L}\tan{\theta_{W}}$.
For the corresponding values for $\beta=\frac{1}{\sqrt{3}}$ and 
$\gamma=\frac{1}{\sqrt{6}}$ we have   
\begin{align}
g_{X} = \frac{g_{L} \tan{\theta_{W}}}{\sqrt{1 - \frac{1}{2}\tan^{2}{\theta_{W}}}}    
\end{align}
Taking into account these choices, we obtain: 
\begin{table}[H]
\begin{center}
\caption{$Z'$ chiral charges for the SM leptons when
embedded in $S_L^3$. Here, $\theta_{W}$ denotes the electroweak mixing angle,  $g_{Y}=g_{L}\tan{\theta_{W}}$, $g_{X} = \frac{g_{L} \tan{\theta_{W}}}{\sqrt{1 - \frac{1}{2}\tan^{2}{\theta_{W}}}}$ and $\psi$ parameterizes the rotation between the additional neutral gauge-boson directions.}
\label{tab:cargasSL3}
\begin{tabular}{| c | c | c |}
\hline
\multicolumn{3}{|c|}{\ \ \  $\ell\subset \mathbf{4}$, $e_R\subset 1$  (as in $S_L^3$) }\\
\hline
SM fields & $g_{Z'} \epsilon_{L}^{Z'}$ & $g_{Z'} \epsilon_{R}^{Z'}$ \\ \hline
$\nu_e$ & $ g_Y\left( -\frac{1}{2 \sqrt{2}}\frac{g_{L}}{g_{X}}-\frac{3}{4\sqrt{2}}\frac{g_{X}}{g_{L}}\right)\sin{\psi}$ & 0  \\ \hline
$e$ & $ g_Y\left( -\frac{1}{2 \sqrt{2}}\frac{g_{L}}{g_{X}}-\frac{3}{4\sqrt{2}}\frac{g_{X}}{g_{L}}\right)\sin{\psi}$ & $ \frac{1}{\sqrt{2}}g_{X}\tan{\theta_{W}}\sin{\psi} $ \\ \hline
\hline
\end{tabular}
\label{tab:c1}
\end{center}
\end{table}

\begin{table}[H]
\begin{center}
\caption{$Z'$ chiral charges for the SM leptons when
embedded in $S_L^5$. Here, $\theta_{W}$ is the electroweak mixing angle, $g_{Y}=g_{L}\tan{\theta_{W}}$  and $g_{X} = \frac{g_{L} \tan{\theta_{W}}}{\sqrt{1 - \frac{1}{2}\tan^{2}{\theta_{W}}}}$.}
\begin{tabular}{| c | c | c |}
\hline
\multicolumn{3}{|c|}{\ \ \  $\ell\subset \overline{\mathbf{4}}$, $e_R\subset 1$  (as in $S_L^5$) }\\
\hline
SM fields & $g_{Z'} \epsilon_{L}^{Z'}$ & $g_{Z'} \epsilon_{R}^{Z'}$ \\ \hline
$\nu_e$ & $ g_Y\left( -\frac{1}{2 \sqrt{2}}\frac{g_{L}}{g_{X}}-\frac{1}{4\sqrt{2}}\frac{g_{X}}{g_{L}}\right)\sin{\psi}$ & 0  \\ \hline
$e$ & $ g_Y\left( -\frac{1}{2 \sqrt{2}}\frac{g_{L}}{g_{X}}-\frac{1}{4\sqrt{2}}\frac{g_{X}}{g_{L}}\right)\sin{\psi}$ & $ \frac{1}{\sqrt{2}}g_{X}\tan{\theta_{W}}\sin{\psi} $ \\ \hline
\hline
\end{tabular}
\label{tab:c2}
\end{center}
\end{table}

\begin{table}[H]
\begin{center}
\caption{$Z'$ chiral charges for the SM quarks when
embedded in $S_Q^1$. Here, $\theta_{W}$ denotes the electroweak mixing angle,  $g_{Y}=g_{L}\tan{\theta_{W}}$, $g_{X} = \frac{g_{L} \tan{\theta_{W}}}{\sqrt{1 - \frac{1}{2}\tan^{2}{\theta_{W}}}}$ and $\psi$ parameterizes the rotation between the additional neutral gauge-boson directions.}
\begin{tabular}{| c | c | c |}
\hline
\multicolumn{3}{|c|}{\ \ \  $q\subset \mathbf{4}$, $u_R, d_R\subset 1$  (as in $S_Q^1$) }\\
\hline
SM fields & $g_{Z'} \epsilon_{L}^{Z'}$ & $g_{Z'} \epsilon_{R}^{Z'}$ \\ \hline
$u$ & $ g_Y\left( -\frac{1}{2 \sqrt{2}}\frac{g_{L}}{g_{X}}-\frac{1}{12\sqrt{2}}\frac{g_{X}}{g_{L}}\right)\sin{\psi}$ & $-\frac{2}{3\sqrt{2}}g_{X}\tan{\theta_{W}}\sin{\psi}$  \\ \hline
$d$ & $ g_Y\left( -\frac{1}{2 \sqrt{2}}\frac{g_{L}}{g_{X}}-\frac{1}{12\sqrt{2}}\frac{g_{X}}{g_{L}}\right)\sin{\psi}$ & $ \frac{1}{3\sqrt{2}}g_{X}\tan{\theta_{W}}\sin{\psi} $ \\ \hline
\hline
\end{tabular}
\label{tab:c3}
\end{center}
\end{table}

\begin{table}[H]
\begin{center}
\caption{$Z'$ chiral charges of the SM quarks when they are embedded in $S_Q^2$. Here, $\theta_{W}$ denotes the electroweak mixing angle,  $g_{Y}=g_{L}\tan{\theta_{W}}$, $g_{X} = \frac{g_{L} \tan{\theta_{W}}}{\sqrt{1 - \frac{1}{2}\tan^{2}{\theta_{W}}}}$ and $\psi$ parameterizes the rotation between the additional neutral gauge-boson directions.}
\label{tab:cargasSQ2}
\begin{tabular}{| c | c | c |}
\hline
\multicolumn{3}{|c|}{\ \ \  $q\subset \overline{\textbf{4}}$, $u_R,d_R\subset 1$  (as in $S_Q^2$) }\\
\hline
SM fields & $g_{Z'} \epsilon_{L}^{Z'}$ & $g_{Z'} \epsilon_{R}^{Z'}$ \\ \hline
$u$ & $ g_Y\left( -\frac{1}{2 \sqrt{2}}\frac{g_{L}}{g_{X}}+\frac{5}{12\sqrt{2}}\frac{g_{X}}{g_{L}}\right)\sin{\psi}$ & $-\frac{2}{3\sqrt{2}}g_{X}\tan{\theta_{W}}\sin{\psi}$  \\ \hline
$d$ & $ g_Y\left( -\frac{1}{2 \sqrt{2}}\frac{g_{L}}{g_{X}}+\frac{5}{12\sqrt{2}}\frac{g_{X}}{g_{L}}\right)\sin{\psi}$ & $ \frac{1}{3\sqrt{2}}g_{X}\tan{\theta_{W}}\sin{\psi} $ \\ \hline
\hline
\end{tabular}
\label{tab:c4}
\end{center}
\end{table}

\subsection{Configuration  \texorpdfstring{$b=1,c=-2$}{}}

\noindent Starting from the Eq. (\ref{qem2}), for the $b=1$ and $c=-2$ configuration, we have that the values of $\beta$ and $\gamma$ are: $\beta= \frac{1}{\sqrt{3}}$ and $\gamma= -\frac{2}{\sqrt{6}}$ respectively. With this and considering that $\tilde\alpha_{L}=\frac{1}{g_{L}}$ of Eq.\eqref{eq:alfa}, the Eq.\eqref{eq:zizq} can be rewritten as:

\setlength{\arraycolsep}{1.4pt}
\begin{align}
 & g_{Z'} \epsilon_{L}^{Z'}=\frac{g_{L}}{2\sqrt{2}}
\left(
\begin{array}{cccc}
1 & 0 & 0 & 0 \\
 0 & 1 & 0 & 0 \\
 0 & 0 & -1 & 0 \\
 0 & 0 & 0 & -1  \\
\end{array}
\right)\cos{\psi}
+ g_{Y}\left(
\begin{array}{cccc}
\frac{g_{X}}{g_{L}}X_{L} & 0 & 0 &
   0 \\
 0 & \frac{g_{X}}{g_{L}}X_{L} & 0 &
   0 \\
 0 & 0 & \frac{1}{2}\frac{g_{L}}{g_{X}}+\frac{g_{X}}{g_{L}}X_{L} & 0 \\
 0 & 0 & 0 & -\frac{1}{2}\frac{g_{L}}{g_{X}}+\frac{g_{X}}{g_{L}}X_{L}
\end{array}
\right)\sin{\psi}
\end{align}

From Eq.~\eqref{eq:zder}, the right-handed chiral charge of $Z'$ is written as:
\begin{align}
g_{Z'} \epsilon_{R}^{Z'}=g_{X}X_{R}\tan{\theta_{W}}\sin{\psi},
\end{align}
\noindent here $g_{Y}=g_{L}\tan{\theta_{W}}$.
The results presented in the following two tables apply to both the fundamental $\mathbf{4}$ and antifundamental $\overline{\mathbf{4}}$ representations of $SU(4)_L$, since the corresponding electroweak charges are identical in both cases.

\begin{table}[H]
\begin{center}
\caption{$Z'$ chiral charges for the SM leptons when
embedded in $S_L^{\prime 3}$ or  $S_L^{\prime 4}$. Here, $\theta_{W}$ is the electroweak mixing angle.}
\label{tab:cargaSL'3}
\begin{tabular}{| c | c | c |}
\hline
\multicolumn{3}{|c|}{\ \ \  $\ell\subset \textbf{4} \ \text{or}\  \overline{\textbf{4}}$, $e_R\subset 1$  (as in $S_L^{\prime 3}$ or  $S_L^{\prime 4}$) }\\
\hline
SM fields & $g_{Z'} \epsilon_{L}^{Z'}$ & $g_{Z'} \epsilon_{R}^{Z'}$ \\ \hline
$\nu_e$ & $\frac{g_{L}}{2\sqrt{2}}\cos{\psi}-\tfrac{1}{2} g_Y\frac{g_{X}}{g_{L}}\sin{\psi}$ & 0  \\ \hline
$e$ & $\frac{g_{L}}{2\sqrt{2}}\cos{\psi}-\tfrac{1}{2} g_Y\frac{g_{X}}{g_{L}}\sin{\psi}$ & $g_{X}\tan{\theta_{W}}\sin{\psi} $ \\ \hline
\hline
\end{tabular}
\label{tab:c5}
\end{center}
\end{table}

\begin{table}[H]
\begin{center}
\caption{$Z'$ chiral charges for the SM quarks when
embedded in $S_Q^{\prime 1}$ or $S_Q^{\prime 2}$. Here, $\theta_{W}$ is the electroweak mixing angle.}
\begin{tabular}{| c | c | c |}
\hline
\multicolumn{3}{|c|}{\ \ \  $q\subset \textbf{4} \ \text{or}\ \overline{\textbf{4}}$, $u_R, d_R\subset 1$  (as in $S_Q^{\prime 1}$ or $S_Q^{\prime 2}$) }\\
\hline
SM fields & $g_{Z'} \epsilon_{L}^{Z'}$ & $g_{Z'} \epsilon_{R}^{Z'}$ \\ \hline
$u$ & $ \frac{g_{L}}{2\sqrt{2}}\cos{\psi}+\tfrac{1}{6} g_Y\frac{g_{X}}{g_{L}}\sin{\psi}$ & $-\frac{2}{3}g_{X}\tan{\theta_{W}}\sin{\psi}$  \\ \hline
$d$ & $ \frac{g_{L}}{2\sqrt{2}}\cos{\psi}+\tfrac{1}{6} g_Y\frac{g_{X}}{g_{L}}\sin{\psi}$ & $ \frac{1}{3}g_{X}\tan{\theta_{W}}\sin{\psi} $ \\ \hline
\hline
\end{tabular}
\label{tab:cargaSQ'2}
\end{center}
\end{table}

\bibliographystyle{apsrev4-2}
\bibliography{references}

\end{document}